\documentclass[dvips]{article}
\newcommand{\doublespacing}{\let\CS=\@currsize\renewcommand{\baselinesstrech}
{2.0}\tiny\CS}
\begin{document}

\textwidth 16cm
\newcommand{\bd}{\begin{document}}
\newcommand{\ed}{\end{document}}
\newcommand{\bc}{\begin{center}}
\newcommand{\ec}{\end{center}}
\newcommand{\bfr}{\begin{flushright}}
\newcommand{\efr}{\end{flushright}}
\newcommand{\lt}{\left}
\newcommand{\rt}{\right}
\newcommand{\vs}{\vspace}
\newcommand{\hs}{\hspace}
\newcommand{\beq}{\begin{equation}}
\newcommand{\eeq}{\end{equation}}
\newcommand{\lb}{\linebreak}
\newcommand{\pb}{\pagebreak}
\newcommand{\mb}{\makebox}
\newcommand{\fb}{\framebox}
\newcommand{\mc}{\multicolumn}
\newcommand{\ben}{\begin{enumerate}}
\newcommand{\een}{\end{enumerate}}
\newcommand{\bit}{\begin{itemize}}
\newcommand{\eit}{\end{itemize}}
\newcommand{\ol}{\overline}
\newcommand{\un}{\underline}
\newcommand{\lefq}{\lefteqn}
\newcommand{\ba}{\begin{array}}
\newcommand{\ea}{\end{array}}
\newcommand{\beqa}{\begin{eqnarray}}
\newcommand{\eeqa}{\end{eqnarray}}
\newcommand{\beqas}{\begin{eqnarray*}}
\newcommand{\eeqas}{\end{eqnarray*}}
\newcommand{\bfg}{\begin{figure}}
\newcommand{\efg}{\end{figure}}
\newcommand{\bds}{\begin{displaymath}}
\newcommand{\eds}{\end{displaymath}}
\newcommand{\btb}{\begin{tabbing}}
\newcommand{\etb}{\end{tabbing}}
\newcommand{\para}{\parallel}
\newcommand{\pad}{\partial}
\newcommand{\nn}{\nonumber}
\newcommand{\la}{\leftarrow}
\newcommand{\ra}{\rightarrow}
\newcommand{\lgla}{\longleftarrow}
\newcommand{\lgra}{\longrightarrow}
\newcommand{\La}{\Leftarrow}\newcommand{\Ra}{\Rightarrow}
\newcommand{\Lra}{\Leftrightarrow}
\newcommand{\Lgla}{\Longleftarrow}
\newcommand{\Lgra}{\Longrightarrow}
\newcommand{\bm}{\boldmath}
\newcommand{\lan}{\langle}
\newcommand{\ran}{\rangle}
\renewcommand{\a}{\alpha}
\renewcommand{\b}{\beta}
\newcommand{\g}{\gamma}
\newcommand{\G}{\Gamma}
\renewcommand{\d}{\delta}
\newcommand{\eps}{\epsilon}
\newcommand{\Th}{\Theta}
\newcommand{\s}{\sigma}
\newcommand{\lam}{\lambda}
\newcommand{\D}{\Delta}
\newcommand{\vare}{\varepsilon}
\newcommand{\pr}{\prime}
\newcommand{\ro}{\rho}
\newcommand{\nab}{\nabla}
\newcommand{\m}{\mu}
\newcommand{\n}{\nu}
\newcommand{\Sg}{\Sigma}
\newcommand{\p}{\pi}
\newcommand{\R}{I\!\!R}
\newcommand{\om}{\omega}
\newcommand{\Om}{\Omega}
\newcommand{\ze}{\zeta}
\newcommand{\vart}{\vartheta}
\newcommand{\tri}{\triangle}
\newcommand{\f}{\frac}
\newcommand{\iny}{\infty}
\newcommand{\pro}{\propto}

\bc {\huge \bf Reply to Comment on "Non Hermitian Quantum Mechanics with Minimal Length Uncertainty", arXiv:0908.2341}
\ec
\vs{2cm}

\bc {\ T.K. Jana {\footnote{email : tapasisi@gmail.com} and P. Roy{\footnote{email : pinaki@isical.ac.in}}\\
Physics \& Applied Mathematics Unit \\
Indian Statistical Institute \\
Kolkata - 700 108, India.}} \ec \vspace{1.5cm}

%\bc {\large {\un{Abstract}}} \ec
{{\bf Abstract} : It is shown that the results of ref \cite{jana} are consistent.}\\\\
\vspace{.25cm}

In ref \cite{jana} Swanson model with the following Hamiltonian was considered
 
\beq
H = \omega a^\dagger a + \lam a^2 + \d{a^\dagger}^2 + \f{\om}{2}\label{swan1}
\eeq
where $\lam\neq \d$ are real numbers and $a,a^\dagger$ are harmonic oscillator annihilation and creation operators given by \footnote{In the expression of $a$ and $a^\dagger$ in ref \cite{jana} $m$ inside the parenthesis was omitted due to typograhical error.} :
\beq
a = \f{1}{\sqrt{2m\hbar \omega}}\left({\hat p}-i\omega m{\hat x}\right),~~~~a^\dagger = \f{1}{\sqrt{2m\hbar \omega}}\left({\hat p}+i\omega m{\hat x}\right)
\label{aad}\eeq
Using Eq.(\ref{aad}) the Hamiltonian (\ref{swan1}) can be written as
\beq
\frac{1}{2m\hbar \omega}\left[(\omega+\lambda+\delta)\hat p^2+im\omega(\delta-\lambda-\omega)\hat p \hat x +im\omega(\delta-\lambda+\omega)\hat x\hat p+m^2\omega^2(\omega-\lambda-\delta)\hat x^2 \right]+\frac{\omega}{2}
\label{swan2}\eeq
Now following ref \cite{bagchi} we put  
\beq
m=\hbar=1,~\lambda=-\delta ~and~\mu=\delta-\lambda \label{con1}
\eeq
and obtain from Eq.(\ref{swan2}) 
\beq
H=\frac{\hat p^2}{2}+\frac{\omega^2 \hat x^2}{2}+i\frac{\mu}{2}\left\{\hat x,\hat p\right\}+i\frac{\omega}{2}\left[\hat x,\hat p\right]+\frac{\omega}{2}
\label{swan3}\eeq
On the other hand the Hamitonian used in ref \cite{bagchi} is given by
\beq
H_{BF}(\hat x,\hat p)=\frac{\hat p^2}{2}+\frac{\omega^2 \hat x^2}{2}+i\mu\left\{\hat x,\hat p\right\}\label{b1}
\eeq
It can be seen that the Hamiltonians (\ref{swan3}) and (\ref{b1}) are not identical as claimed in ref \cite{bagchi}. Also even for $\b=0$ the Hamiltonians (\ref{swan3}) and (\ref{b1}) are different.
\vspace{.25cm}

Now in the undeformed case i.e, for $\b=0$ the Hamiltonian (\ref{swan3}) becomes
\beq
H_0=Q\frac{\partial^2}{\partial p^2}+Rp\frac{\partial}{\partial p}+Sp^2+T\label{swan4}
\eeq
where $Q=-\frac{m\hbar \omega}{2}(\omega-\lambda-\delta),~	R=(\lambda-\delta),~S=\frac{\omega+\lambda+\delta}{2m\hbar \omega},~T=\frac{\lambda-\delta}{2}$ and it's conjugate is
\beq
H_0^\dagger=Q\frac{\partial^2}{\partial p^2}-Rp\frac{\partial}{\partial p}+Sp^2-T
\eeq
We recall that the metric operator for the Swanson Hamiltonian (\ref{swan3}) was found in ref \cite{jana} as
\beq
\eta=(1+\beta p^2)^{\frac{(\delta-\lambda)}{m\hbar \omega(\omega-\lambda-\delta)\beta}}\label{eta}
\eeq
Now the metric operator for the undeformed Hamiltonian (\ref{swan4}) can be found to be
\beq
\eta_0=\lim_{\b\rightarrow 0}\eta=e^{\frac{(\delta-\lambda)p^2}{m\hbar \omega(\omega-\lambda-\delta)}}
\eeq
and if (\ref{con1}) is used then it reduces to
\beq
\eta_0=e^{\f{\mu p^2}{\omega^2}}
\eeq
It can be easily checked that
\beq
\eta_0 H_0 \eta_0^{-1}=H_0^\dagger
\eeq

Finally we refer to twofold non Hermiticity mentioned in ref \cite{bagchi}. In ref \cite{jana}, quantum mechanics with minimal length has been described in sec 2 and Eq.(4) of ref \cite{jana} describes the scalar product for such models. We maintain that the Hamiltonians in Eq.(5) and Eq.(13) of ref \cite{jana} are Hermitian when $\lam=0$ and $\lam=\delta$ respectively within the framework of sec 2 of ref \cite{jana} i.e, no non Hermiticity is introduced when the transition $(x,p)\rightarrow ({\hat x},{\hat p})$ is made.

\ed
\begin{thebibliography}{20}
\bibitem{jana} T.K. Jana and P. Roy, Non-Hermitian Quantum Mechanics
with Minimal Length Uncertainty, SIGMA 5 (2009), 083, 7 pages; arXiv:0908.1755.\footnote[4]{After this paper was submitted for publication a paper \cite{bagchi1} dealing topics similar to \cite{jana} appeared. Unfortunately this paper escaped our attention during the proof correction of ref \cite{jana}.} 
\bibitem{bagchi} B. Bagchi and A. Fring, SIGMA 5 (2009), 089, 2 pages. 
\bibitem{bagchi1} B. Bagchi and A. Fring, preprint arXiv:0907.5354.
\end{thebibliography}
